\documentclass[prb,12pt,epsfig]{revtex4}
\usepackage[english]{babel}
\usepackage[intlimits]{amsmath}
\usepackage[dvips]{graphicx}
\usepackage{epsfig}

\begin{document}

\begin{titlepage}
\
\title{Effect of  Inhomogeneity in Translocation of Polymers through Nanopores}
\author{Stanislav Kotsev and Anatoly B. Kolomeisky}

\affiliation{Department of Chemistry, Rice University, Houston, TX 77005-1892 USA}

\begin{abstract}
The motion of polymers with inhomogeneous  structure through nanopores  is discussed theoretically. Specifically, we consider the translocation dynamics of polymers consisting of double-stranded and single-stranded blocks. Since only the single-stranded chain can go through the nanopore the double-stranded segment has to unzip before the translocation.  Utilizing a simple analytical model, translocation times  are calculated explicitly for  different polymer orientations, i.e., when the single-stranded block enters the pore first and when the double-stranded segment is a leading one. The dependence of the translocation dynamics on external fields, energy of interaction in the double-stranded segment, size of the polymer and the fraction of double-stranded monomers is analyzed. It is found that the order of entrance into the pore has a significant effect on the translocation dynamics. The theoretical results are discussed using free-energy landscape arguments.
\end{abstract}

\maketitle

\end{titlepage}

\section{Introduction}

Translocation of polymer molecules through nanopores plays a critical role in numerous  chemical and biological  processes.\cite{lodish_book,meller_review} Examples include the motion of DNA and RNA molecules across nuclear pores, injection of viral DNA into cells, gene swapping and protein transport through membranes. Because of its importance for understanding fundamental processes in biology\cite{meller_review} and potential technological applications\cite{deamer02} the polymer translocation has become a subject of intensive experimental,\cite{kasianowicz96,henrickson00,meller00,meller02,sauer-budge03,bates03,mathe04,wang04,storm05,mathe05,mathe06,butler06} theoretical\cite{sung96,lubensky99,muthukumar99,muthukumar02,ambjornsson02,slonkina03,flomenbom03,metzler03,kafri04,lakatos05,bundschuh05} and numerical\cite{mathe05,chuang01,chern01,zandi03,milchev04,rabin05,muthukumar06,matysiak06} studies.

In {\it in vitro} experiments the polymer chain threads through the single protein channel, like the bacterial $\alpha$-hemolysin, that is positioned in a membrane.\cite{kasianowicz96,henrickson00,meller00,meller02,sauer-budge03,bates03,mathe04,wang04,mathe05,mathe06,butler06} For the polymer moving through the pore the number of available configurations decreases, creating an effective entropic barrier for translocation. To overcome this barrier the polymers, which are typically DNA or RNA molecules with charged bases, are driven across the membrane  with the help of external electric fields. When the polymer is in the pore the membrane current, produced mainly by small ions present in the system, decreases significantly. The translocation properties of every polymer molecule is specified by measured translocation times and blockade currents. Because the narrowest part of $\alpha$-hemolysin channel is less than 2nm in diameter,\cite{meller_review,meller00,meller02,sauer-budge03,bates03,mathe04,wang04,mathe05,mathe06,butler06} only single-stranded DNA or RNA molecules can pass through the channel. In addition, the translocation  for not very long DNA and RNA molecules is typically much faster than other biologically important processes, such as inter- and intra-molecular interactions.\cite{meller_review} Explicit measurements\cite{mathe04} showed that for the range of voltages from 30 to 150 mV the translocation time for the single-stranded DNA with 26 bases is less than 0.01 of the time needed to unzip a hairpin with 10 bases under the same conditions.  These observations allow to utilize the nanopore experiments to test the structure, chemical states and interactions of single polymer molecules. For example, in Ref.\cite{sauer-budge03} single-molecule dynamics of unzipping of double-stranded DNA has been investigated by pulling one of the strands through  $\alpha$-hemolysin pore at different voltages and temperatures. Other recent experiments\cite{mathe05} also show an orientation asymmetry in the translocation of single-stranded DNA molecules, which is due to  complex interactions of DNA nucleotides with the protein nanopore.

Theoretically, the translocation dynamics is generally viewed as an effective one-dimensional diffusion  in a free-energy landscape that is biased by en external driving field.\cite{sung96,lubensky99,muthukumar99,muthukumar02,ambjornsson02,slonkina03,flomenbom03,metzler03,kafri04,lakatos05,bundschuh05} The translocation times are associated with  mean first-passage times that can be calculated from the free-energy potentials.\cite{muthukumar99,slonkina03} Most of the theoretical works consider only  homogeneous polymers, although several nanopore  experiments\cite{sauer-budge03,mathe04,wang04,butler06} indicate that in the real biological systems inhomogeneities in the structure and interactions of translocating polymers with other molecules  might have a significant effect on the overall dynamics. Muthukumar\cite{muthukumar02} theoretically discussed some effects of the sequence on DNA translocation through nanopores by considering the motion of  diblock copolymers. It was found that for the weak driving forces the passage time depends on which polymer end begins the translocation. However no explanation for this phenomenon has been given. The effect of sequence heterogeneity in polymer translocation was also studied in Ref.\cite{kafri04}  where it was shown that the heterogeneity might lead to anomalous dynamics at some conditions.

In this paper we investigate the translocation dynamics of inhomogeneous polymers using simple analytical model and free-energy landscape arguments. Specifically, we consider the motion of polymers consisting of two blocks. One is a single-stranded chain and  another one is a double stranded segment. Translocation times at different external conditions are calculated explicitly for two cases: 1) when the single-stranded part moves through the nanopore first and then the double-stranded block unzips and translocates; and 2) when the double-stranded end comes first to the pore, which means that this segment starts to unzip  and then one of the unwounded chains enters the pore followed by the translocation of the single-stranded piece.

The paper is organized as follows. The description of the model and our theoretical method  are presented in Sec. II, while the results and predictions are discussed in Sec. III. Section IV summarizes and concludes our theoretical analysis. Some details of the mathematical calculations are given in Appendix.

\section{Our Model}

Let us consider a translocation through the pore of the inhomogeneous polymer molecule with $N$ monomers as shown in Fig. 1. Each monomer has a charge $q$. One part of the molecule is a single-stranded segment, while the other part is made of double-stranded segment. The composition of the polymer is specified by a parameter $a$ ($0 \le a \le 1$), i.e.,  $Na$ monomers belong to the double-stranded block with a bond energy per monomer $\varepsilon$. We assume that the diameter of the nanopore is so small that only the single-stranded polymer chain can go through the channel. If the polymer comes to the entrance of the pore with the double-stranded segment first, then it starts to unzip and only the longest chain goes through the nanopore - see Fig. 1.  It is also assumed that the length of the nanopore is small so that no more than one monomer can be found inside. It is a good approximation for the polymers whose contour length is much larger than the nanopore length. Note, however, that the length of the nanopore is an important factor in the translocation dynamics of real polymers.\cite{meller_review,slonkina03,flomenbom03} In addition, we discuss only the translocation process after the polymer entered into the pore, and it is assumed that the molecule that entered into the pore will not exit back. The problem of searching for the nanopore is beyond the scope of this work.

To overcome the entropic barrier the polymer is driven across the channel by an external electric field of strength $V$. In the model the hydrodynamic force is neglected since theoretical calculations show that it is smaller than the electric force in the {\it in vitro} experiments.\cite{lubensky99,slonkina03}  We define $P_{k}(t)$ as a probability to find $k$ monomers already translocated at time $t$. The dynamics of the system at all times is governed by a Master equation,
\begin{equation}
\frac{dP_{k}(t)}{dt}=u_{k-1}P_{k-1}(t)+w_{k+1}P_{k+1}(t)-(u_{k}+w_{k}) P_{k}(t),
\end{equation}
where $u_{k}$ ($w_{k}$) is the rate to move the polymer molecule by one monomer in the forward (backward) direction. The rates are related by the detailed balance condition,
\begin{equation}\label{detailed_balance}
\frac{u_{k}} {w_{k+1}}=\exp[-\beta(F_{k+1}-F_{k})]=\exp[-\beta \Delta F_{k}],
\end{equation}
where $F_{k}$ is a free energy of the polymer with $k$ monomers already moved across the nanopore and $\beta=1/k_{B}T$. It is also assumed that if the free-energy difference of moving the monomer forward or backward is  zero, i.e. $\Delta F_{k}=0$ for all $k$, the forward and backward rates are the same.  

The free energy of the translocating polymer,  shown in Fig. 1 has three terms, namely, entropic, electrostatic and the hybridization energy,
\begin{equation}
F_{k}=F_{k,entr} +F_{k,elec} + F_{k,hyb}.
\end{equation} 
However for realistic values of external fields and unzipping energies the entropic contribution can be neglected.\cite{muthukumar99,slonkina03} Specific values of the free energy depend on the entrance order into the nanopore. For situations when the single-stranded segment is entering first the free energy is given by
\begin{equation}\label{free_energy1}
F_{k}=\left\{ \begin{array}{cc}
              -kqV - \varepsilon Na, & \mbox{ for } k <N(1-a); \\
              -k(qV - \varepsilon) - \varepsilon N, & \mbox{ for } k >N(1-a), 
              \end{array} \right.
\end{equation}
where $q$ is an effective charge per monomer, and $\varepsilon$ is a hybridization energy per monomer. If the double-stranded segment is first at the entrance into the nanopore, then the free energy of the system is different. It also a time-dependent function. Before full unzipping of the double-stranded segment we have
\begin{equation}\label{free_energy2}
F_{k}=-k(qV-\varepsilon) -\varepsilon Na, \quad \mbox{ for } k< Na.
\end{equation}
We assume that the process of full unzipping is irreversible, and the free-energy potential changes after that moment. It becomes equal to $F_{k}=-kqV$ for all values of $k$. The analysis of Eqs. (\ref{free_energy1}) and (\ref{free_energy2}) indicate that at $\varepsilon=qV$ the free-energy surfaces change the behavior. As we will see later, it has a strong effect on the translocation dynamics.

Dynamic properties of translocating polymers depend on the forward and backward rates. However, the detailed balance condition (\ref{detailed_balance}) specifies only the ratio of the rates. In this work for simplicity we assume that the free energy change influences only the forward rates. The rates are given by
\begin{equation}\label{uk}
u_{k}=w' \exp[\beta qV], \quad w_{k}=w',
\end{equation}
for moving the single-stranded segment across the pore, and for transporting the single chain after unzipping the double-stranded segment,
\begin{equation}\label{wk}
u_{k}=w \exp[\beta (qV-\varepsilon)], \quad w_{k}=w.
\end{equation}
More general cases when the free energy change can be distributed between the forward and backward rates can be easily accommodated in our theoretical approach, however, qualitatively it does not change the results.   In Eqs. (\ref{uk},\ref{wk}) parameters $w$ and $w'$ are attempts frequencies, and they describe the motion in the absence of external forces and hybridization. Note, that in our simplified model the asymmetric interactions between the pore and the polymer\cite{mathe05} are not considered, although they can be easily incorporated. However, since it is found that the unzipping controls the translocation\cite{mathe04}, we believe that asymmetric interactions are less important in our case.

In this theoretical approach both forward and backward transitions in the translocation process are taken into account. These transitions are easy to understand for the motion of the single-stranded segment. However, there is some ambiguity for the double-stranded region. We assume here that the forward step, when the doubles-stranded region moves, corresponds to unzipping that is followed  by the forward translocation of one monomer. Similarly, the backward step describes the reverse translocation followed by zipping. Obviously, other processes can take place in the system but the model describes the most probable events of translocation of diblock copolymer. Recent theoretical calculations show that at low external fields the backward transitions out of the pore might become an important factor.\cite{lakatos05}

The dynamics of the polymer moving through the nanopore is described by its translocation time. We define $\tau_{1}$ as the translocation time of the polymer with the single-stranded segment entering first into the channel. Correspondingly, $\tau_{2}$ is the translocation time when the single-stranded segment moves into the pore after unzipping and translocation of the double-stranded block (see Fig. 1). Translocation times are calculated exactly for given parameters using the standard expression for mean first-passage times.\cite{vanKampen,pury03} Then to compare two different translocation mechanisms we introduce a dimensionless ratio of passage times,
\begin{equation}
r=\tau_{1}/\tau_{2}.
\end{equation}

\section{Results and Discussion}

The passage times are computed exactly  for the  different sets of parameters. Explicit expressions are quite bulky and presented in Appendix. The analysis of  formulas  (\ref{time_double_first},\ref{time_single_first}) suggests the translocation times  can be written as a sum of three terms,
\begin{equation}\label{three_times}
\tau_{i}=\tau(s)+\tau(d)+\tau_{i}(corr), \quad i=1,2,
\end{equation}
where $\tau(s)$ is the time to move only the single-stranded segment of the polymer and $\tau(d)$ corresponds to the time to thread only the segment that was double-stranded before the translocation. Obviously, these two contributions are independent of the orientation of the translocating polymer. The last term in Eq. (\ref{three_times}) is a correlation time that reflects backward trajectories when some part of the polymer already unzipped in the previous steps might zip again. It depends on the orientation of the polymer in the nanopore, and it plays a critical role in the translocation dynamics.

 Translocation dynamics depends on the external voltage $V$, hybridization energy $\varepsilon$, the size $N$ of the polymer molecule  and the fraction $a$ of the double-stranded monomers. Exact analytical formulas allow us to consider  these effects in detail. In the limit of high external voltages the translocation times in both orientations  become equal, and they exponentially  approach  zero as follows,
\begin{equation}
 \tau_{1}=\tau_{2}\simeq \frac{N(1-a)}{w'}\exp(-\beta qV) + \frac{Na}{w}\exp[\beta(\varepsilon-qV)] \rightarrow 0.
\end{equation}
Because of the very strong field, in this regime the probability of backward transitions is very low, and the correlation time is essentially equal to zero. This leads to the dynamics that does not distinguish between the orientation of the translocating polymer.  The situation is very different for the weak external forces, where the explicit formulas (\ref{time_double_first},\ref{time_single_first})  yield,
\begin{equation}
\tau_{1} =\frac{1}{2w'}[N^2(1-a)^2-N(1-a)] +  \frac{N(1-a)[1-\exp(\beta Na \varepsilon)]}{w'[\exp(-\beta\varepsilon)-1]} + A,
\end{equation}
and
\begin{equation}
\tau_{2}=\frac{1}{2 w'}[(1-{a^2}){N^2}-N(1-a)] + A,
\end{equation}
with an auxiliary function $A$ defined as
\begin{eqnarray}
 A & = & \frac{Na}{w} \exp(\beta \varepsilon) +\frac{N(1-a)}{w'} +  \frac{Na\exp(2\beta \varepsilon)}{w[1-\exp(\beta \varepsilon)]}  \nonumber \\
&-& \frac{\exp[\beta \varepsilon(Na+1)]}{w[1-\exp(\beta \varepsilon)]} -\frac{\exp(2\beta \varepsilon)}{w[1-\exp(\beta \varepsilon)]^2}+  \frac{\exp[\beta \varepsilon(Na+1)]}{w[1-\exp(\beta \varepsilon)]^2}.
\end{eqnarray}
Passage times to move a polymer molecule across the channel from two opposite configurations differ at low external fields. These results are illustrated in Fig. 2 where the ratio of translocation times $r$ is plotted as a function of external voltage. At $qV/k_{B}T \ll 1$ the correlation terms in the translocation times differ significantly for the two orientations.  This can be understood by considering the situation when one block is already fully translocated. If the translocated block is the single-stranded one, then the forward motion is energetically unfavored since it require to unzip  double-stranded monomers. As a result, there will be a lot of backward transitions. In the opposite orientation, when the double-stranded region already passed the channel, the situation is different. Here the probability  of forward and backward steps is equal and not so many backward trajectories are observed. As a result we have $\tau_{1} > \tau_{2}$, and it was observed that this inequality generally holds for all parameter's space.

Varying hybridization energy  has also a strong effect on the translocation dynamics. When $\varepsilon=0$ two orientations are indistinguishable (for $w=w'$) and, as expected, $r=\tau_{1}/\tau_{2}=1$. However, for strong interactions in the double-stranded segment it is important which polymer segment enters the nanopore first. From Eqs. (\ref{time_double_first},\ref{time_single_first}) it can be shown that for large $\varepsilon$
\begin{equation}
 \tau_{1} = \frac{\exp[\beta(Na+2)(\varepsilon -qV)]}{w [1-\exp(\beta(\varepsilon -qV)]^2} + \frac{\exp[\beta(\varepsilon -2 qV)] \exp[\beta Na(\varepsilon - qV)]\left(\exp[-\beta N(1-a)qV] -1\right)}{w'[1-\exp(\beta(\varepsilon- qV)][1- \exp(-\beta qV)]},
\end{equation}
and
\begin{equation}
\tau_{2} = \frac{\exp[\beta(Na+2)(\varepsilon-qV)]}{w[1-\exp[\beta(\varepsilon-qV)]^2}.
\end{equation}
Then the ratio of translocation times for $\varepsilon \gg 1$ is given by
\begin{equation}
r=1+ \frac{w (\exp(-\beta qV)-\exp[-\beta(N(1-a)-1)qV])}{w'[1 - \exp(-\beta qV)]}.
\end{equation}
The dependence of $r$ on hybridization energy is plotted in Fig. 3. It can be seen that strong intra-molecular interactions significantly slow down the translocation of polymers with single-stranded segment entering first. This is due to the fact that the slow unzipping allows the single-stranded polymer chain to make many forward and backward threading transitions. It leads to $\tau_{1}$ being typically much larger than $\tau_{2}$ at these conditions. This effect is even stronger for weaker external voltages.

An important parameter in the motion of polymers across the channels is the molecular size $N$.\cite{meller00,slonkina03,flomenbom03} Our theoretical calculations for diblock copolymers with $N \rightarrow \infty$ show that the translocation depends on the relative values of hybridization energy and electrostatic energy. From Eqs. (\ref{time_double_first},\ref{time_single_first}) for $\varepsilon > qV$ the passage times are equal to
\begin{equation}
\tau_{1}=\frac{\exp[\beta(Na+2)(\varepsilon-qV)]}{w[1-\exp[\beta(\varepsilon-qV)]^2} - \frac{\exp[\beta(\varepsilon(Na+1)-qV(Na+2))]}{w'[1-\exp[\beta(\varepsilon-qV)][1-\exp(-\beta qV)]},
\end{equation}
and 
\begin{equation}
\tau_{2}= \frac{\exp[\beta (Na+2)(\varepsilon-qV)]}{w[1-\exp[\beta(\varepsilon-qV)]^2}.
\end{equation}
It is interesting to note that at these conditions both $\tau_{1}$ and $\tau_{2}$ depend on $N$ only through $Na$. This means that the translocation times are completely determined by the size of the double-stranded segment and they do not depend on the size of the single-stranded block. The translocation dynamics changes dramatically for large-size  polymers when the electric field starts to dominate, i.e., when $\varepsilon < qV$. In this case in both orientations the polymer moves through the nanopore with equal times,
\begin{eqnarray}
\tau_{1}=\tau_{2}& = &\frac{N(1-a)\exp(-\beta qV)}{w'} + \frac{Na\exp[\beta(\varepsilon-qV)]}{w} + \frac{N(1-a)\exp(-2\beta qV)}{w'[1-\exp(-\beta qV)} \nonumber \\
& + & \frac {Na\exp[2\beta (\varepsilon-qV)]}{w[1-\exp[\beta(\varepsilon-qV)]}.
\end{eqnarray}
This is exactly the sum of the passage times for two polymer blocks separately, i.e., the correlation term in the translocation time [see Eq. (\ref{three_times})] is negligible at these conditions  when the electric field dominates the passage dynamics. Thus we obtain for $N \gg 1$,
\begin{equation}\label{ratio_limit_highN}
r= \left\{ \begin{array}{lr} 
1, &     \varepsilon <qV; \\
1+ \frac{w\exp[\beta(\varepsilon-2qV)](\exp[\beta(\varepsilon-qV)]-1)}{w'\exp[2\beta(\varepsilon-qV)][1-\exp(-\beta qV)]},  & \varepsilon >qV.
\end{array} \right.   
\end{equation}
The fact that $\varepsilon=qV$ specifies a dynamic phase boundary could also be understood from free-energy potentials - see Eqs. (\ref{free_energy1}) and (\ref{free_energy2}). The ratio of translocation times as a function of polymer size $N$ is shown at Fig. 4. This situation is an example of a dynamic phase transition ($N \rightarrow \infty$) when changing the electrostatic energy contribution  (by modifying $V$ and/or the effective charge $q$) leads to a transition from the orientation-dependent transport to the dynamics that does not discriminate between the polymer orientations. For realistic systems $\varepsilon \simeq$ 2-3 $k_{B}T$, and for typical translocation voltages $V=100$ mV and $q=1e$ we obtain at room temperature $qV  \simeq 4$ $k_{B}T$.  However, the  charge on the monomers can be significantly lower due to screening.\cite{rabin05} This suggests that the predicted dynamic transition might be observed in the nanopore experiments by modifying the voltage or the ionic strength (to change the charge). The main observable feature in this transition is a width of the passage-time distributions. For orientation-dependent dynamics the width is wide  because of possible overlapping of two distinct distributions. At the same time, the width narrows for the dynamics that does not depend on the polymer orientation. It can be suggested that this result possibly can be used as an experimental method of measuring  effective charge $q$ at each monomer. Furthermore, this phenomenon could be important in biological processes. 

Another parameter that can influence the translocation dynamics is the fraction $a$ of the double-stranded monomers. The situations when $a=0$ and $a=1$ are trivial since they correspond to the case of only single-stranded or double-stranded polymers, for which there is no orientation dependence. The general case $0<a<1$ is more complex with a non-monotonous behavior, as can be seen in Fig. 5. The maximal discrimination between two different polymer orientations depends on the hybridization energy and the external voltage.

Our theoretical calculations of the translocation of diblock copolymers predict several unusual orientation-dependent phenomena. This dependence can be explained using free-energy landscape arguments. The polymer molecule {\it before} entering the nanopore has the same free energy in both configurations. Similarly, after moving completely through the channel there is no difference in free energy of the polymer in both orientations. However, when only the part of the molecule translocated the free energy of the system depends on what segment of the polymer entered first into the pore. This indicates that, although the initial and final points on the free-energy landscape are the same, polymers in  different orientations follow  different free-energy pathways,leading to  orientation-dependent phenomena described above.This is similar to a chemical reaction with and without catalyst. The catalyzed reaction is faster because it follows a different free-energy pathway.

\section{Summary and Conclusions}

We developed a simple theoretical model for the translocation through the nanopore  polymer molecules consisting of single-stranded segment and double-stranded segment. Using exact analytical results for the mean first-passage times we show that for some sets of parameters the translocation dynamics of inhomogeneous polymers depends on the orientation of the molecule at the entrance. The translocation for two orientations is different al low external forces, while it disappears at large voltages. Strong hybridization energies also lead to orientation-dependent dynamics. The effect is stronger for large polymer sizes. In addition, we also observed an interesting dynamic phase transition at large $N$ by changing  the relative contributions of external forces and hybridization energy. Our theoretical results are discussed in terms of the free-energy landscape arguments. 

The presented theoretical model of translocation is rather oversimplified. In order to have more realistic description several factors should be included in the analysis. These factors include entropic contributions to free energy that are important at weak external forces, the effect of sequence dependence in rates and hybridization energies, and the effect of folding and unfolding before and after translocation. It will be also very interesting to fully investigate the observed dynamic transition.

\section*{Acknowledgments}

The authors would like to acknowledge the support from the Welch Foundation (Grant No. C-1559), the Alfred P. Sloan Foundation (Grant No. BR-4418), and the U.S. National Science Foundation (Grant No. CHE-0237105). The authors are also grateful to C. Clementi and M. Pasquali for valuable discussions.

\section*{Appendix}

\setcounter{equation}{0}

\renewcommand{\theequation}{\mbox{A\arabic{equation}}}

Translocation times can be calculated using the well-known exact formula for  mean first passage times for a one-dimensional asymmetric discrete random walk.\cite{vanKampen,pury03} For the case when the double-stranded block is first at the entrance into the pore it is given by
\begin{eqnarray}
 \tau_{2} & = & \frac{Na}{w}\exp[\beta(\varepsilon-qV)] +\frac{N(1-a)}{w'}\exp(-\beta qV)+ \frac{\exp[\beta(\varepsilon-qV)]}{w} \sum_{k=1}^{Na-1} (Na-k)\exp[\beta k(\varepsilon-qV)]\nonumber \\
 & + &  \frac{\exp(-\beta qV)}{w'} N(1-a)\sum_{k=1}^{Na}\exp(-\beta kqV)   \nonumber \\
 & + & \frac{\exp(-\beta qV)}{w'}\sum_{k=1}^{N(1-a)-1}[N(1-a)-k] \exp[-\beta(Na+k)qV].
\end{eqnarray}
Direct summation then yields the following final expression,
\begin{eqnarray}\label{time_double_first}
\tau_{2} & = & \frac{Na}{w} \exp[\beta(\varepsilon-qV)] +\frac{N(1-a)}{w'} \exp(-\beta qV) + \frac{Na\exp[2\beta(\varepsilon-qV)]}{w(1-\exp[\beta(\varepsilon-qV)])}\nonumber \\
 & - & \frac{\exp[\beta(Na+1)(\varepsilon-qV)]}{w [1-\exp[\beta(\varepsilon-qV)]}-\frac{\exp[2\beta(\varepsilon-qV)]}{w[1-\exp[\beta(\varepsilon-qV)]^2} + \frac{\exp[\beta(Na+1)(\varepsilon-qV)]} {w [1-\exp[\beta(\varepsilon-qV)]^2} \nonumber \\
& + & N(1-a)\frac{\exp(-2\beta qV)}{w'[1-\exp(-\beta qV)]} -  \frac{\exp[-\beta(N+1)qV]}{w'[1-\exp(-\beta qV)]} -\frac{\exp[-\beta(Na+2)qV]}{w'[1-exp(-\beta qV)]^2}\nonumber \\
& + &\frac{exp[-\beta(N+1)qV]}{w'[1-\exp(-\beta qV)]^2}.
\end{eqnarray}

Similar calculations for the single-stranded segment translocating first produce 
\begin{eqnarray}
\tau_{1}& = &\frac{Na}{w}\exp[\beta(\varepsilon-qV)] +\frac{N(1-a)}{w'}\exp(-\beta qV) + \frac{\exp[\beta(\varepsilon-qV)]}{w}\sum_{k=1}^{Na-1} (Na-k)\exp[\beta k(\varepsilon-qV)] \nonumber \\
& + &\frac{\exp(-\beta qV)}{w'}\sum_{k=1}^{N(1-a)-1} [N(1-a)-k]\exp(-\beta kqV) + \frac{1}{w'}\sum_{k=1}^{Na}\sum_{i=1}^{N(1-a)}\exp[\beta (k\varepsilon-(i+k)qV)],
\end{eqnarray}
that can be written in the following final form after summation of all terms,
\begin{eqnarray}\label{time_single_first}
\tau_{1}&=& \frac{Na}{w}\exp[\beta(\varepsilon-qV)] +\frac{N(1-a)}{w'}\exp(-\beta qV) +\frac{\exp[(Na+2)\beta(\varepsilon-qV)]}{w[1-\exp[\beta(\varepsilon-qV)]^2} \nonumber \\
&+&{\frac{Na}{w}}{\frac{\exp[2\beta(\varepsilon-qV)]}{1-\exp[\beta(\varepsilon-qV)]}}-\frac{\exp[2\beta(\varepsilon-qV)]}{w[1-\exp[\beta(\varepsilon-qV)]^2} \nonumber \\
&+&\exp[-\beta N(1-a)qV]{\frac{\exp(-2\beta qV)}{w'[1-\exp(-\beta qV)]^2}} +{\frac{N(1-a)}{w'}}{\frac{\exp(-2\beta qV)}{1-\exp(-\beta q V)}}\nonumber \\
&-&\frac{\exp(-2\beta qV)}{w'[1-\exp(-\beta qV)]^2}+\frac{\exp[\beta(\varepsilon-2qV)]}{w'[1-\exp[\beta(\varepsilon-qV)]][1-\exp(-\beta qV)]}\nonumber \\
&+&\frac{\exp[\beta Na(\varepsilon-qV)-\beta N(1-a)qV]\exp[\beta(\varepsilon-2qV)]}{w'[1-\exp[\beta(\varepsilon-qV)]][1-\exp(-\beta qV)]}-\frac{\exp[-\beta N(1-a)qV]\exp[\beta(\varepsilon-2qV)]}{w'[1-\exp[\beta(\varepsilon-qV)]][1-\exp(-\beta qV)]} \nonumber \\
&-&\frac{\exp[\beta Na(\varepsilon-qV)]\exp[\beta(\varepsilon-2qV)]}{w'[1-\exp[\beta(\varepsilon-qV)]][1-\exp(-\beta qV)]}.
\end{eqnarray}

\newpage

\noindent {\bf Figure Captions:} \\

\noindent Fig. 1. \quad  Schematic view of moving a diblock copolymer, made of single-stranded and double-stranded segments, across the nanopore. Two different polymer orientations are shown. Only single-stranded chain can travel  through the nanopore. $Na$ is number of monomers in the double-stranded region. a) Translocation with the single-stranded segment entering the nanopore first.  b) Translocation with the double-stranded segment moving first and unzipping. 

\vspace{5mm}

\noindent Fig. 2. \quad Ratio of translocation times as a function of external electric field. The hybridization is $\varepsilon=2$ $k_{B}T$. Solid lines describe  polymers with $N=30$ monomers, while dashed lines are for polymers with $N=60$ monomers. In each set upper curves correspond to $a=0.25$, and lower curves are for $a=0.75$. For all calculations it is assumed that $w=w'$.

\vspace{5mm}

\noindent Fig. 3. \quad  Ratio of translocation times as a function of hybridization energy $\varepsilon$. Polymer molecules with $N=30$ monomers is considered. For solid lines $a=0.75$, while for dashed lines $a=0.25$. In each set upper curves correspond to $qV/k_{B}T=0.05$, and lower curves are for $qV/k_{B}T=0.1$. For all calculations it is assumed that $w=w'$.

\vspace{5mm}

\noindent Fig. 4. \quad Ratio of translocation times as a function of polymer size $N$. The hybridization is $\varepsilon=2$ $k_{B}T$. Solid lines describe the translocation with $qV/k_{B}T=0.1$, and for dashed lines $qV/k_{B}T=0.05$. In each set upper curves correspond to $a=0.25$, middle curves correspond to $a=0.5$, and lower curves are for $a=0.75$. For all calculations it is assumed that $w=w'$.

\vspace{5mm}

\noindent Fig. 5. \quad Ratio of translocation times as a function of the relative size of the double-stranded segment. For calculations it is assumed that $\varepsilon=2$ $k_{B}T$, $qV/k_{B}T=0.1$ and  $w=w'$. The upper curve describes the polymer with $N=60$ monomers, while the lower curve corresponds to $N=30$.

\newpage

\begin{figure}[ht]
\begin{center}
\unitlength 1in
\begin{picture}(4.0,4.0)
  \resizebox{3.375in}{3.375in}{\includegraphics{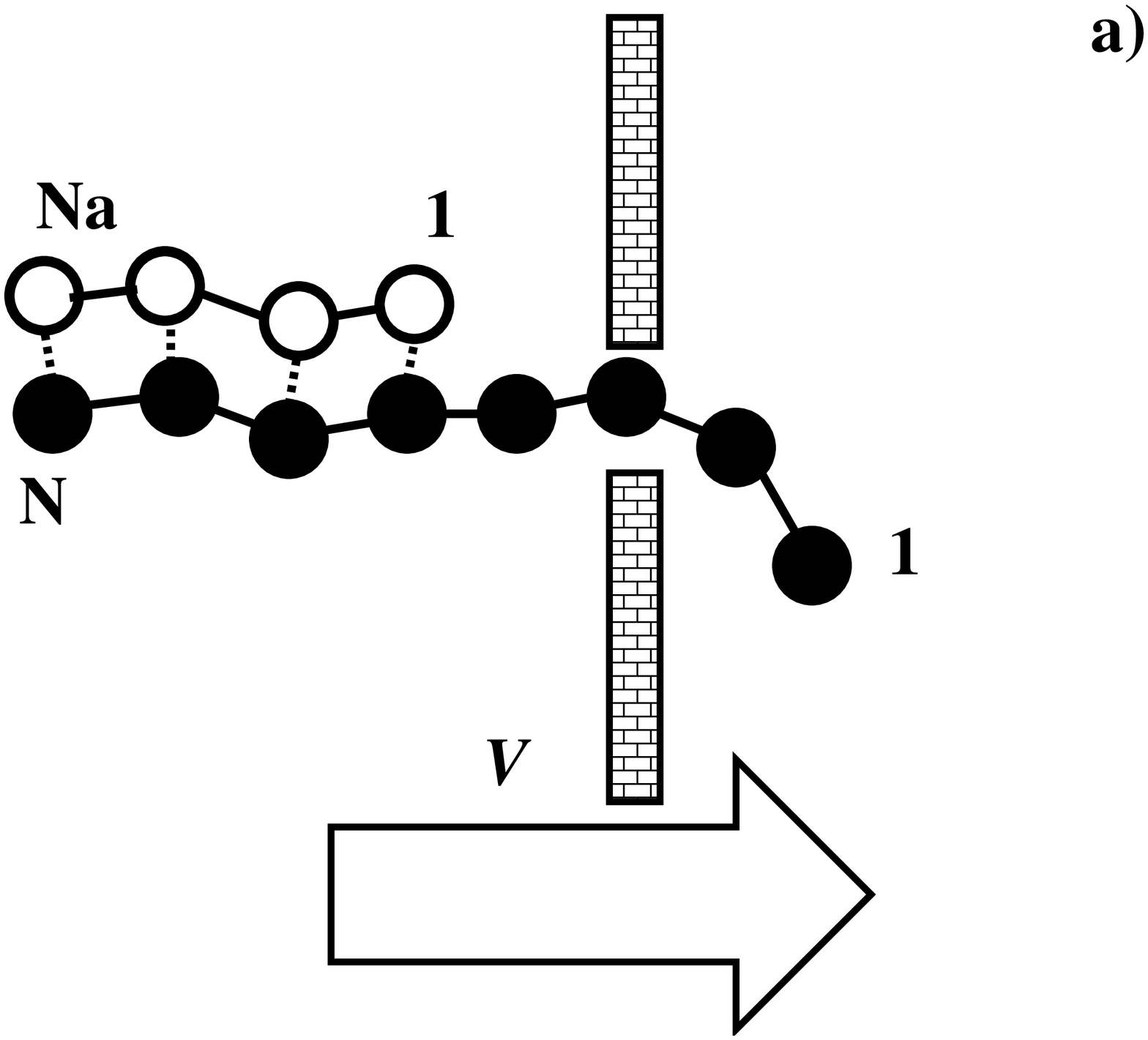}}
\end{picture}
\begin{picture}(4.0,4.0)
  \resizebox{3.375in}{3.375in}{\includegraphics{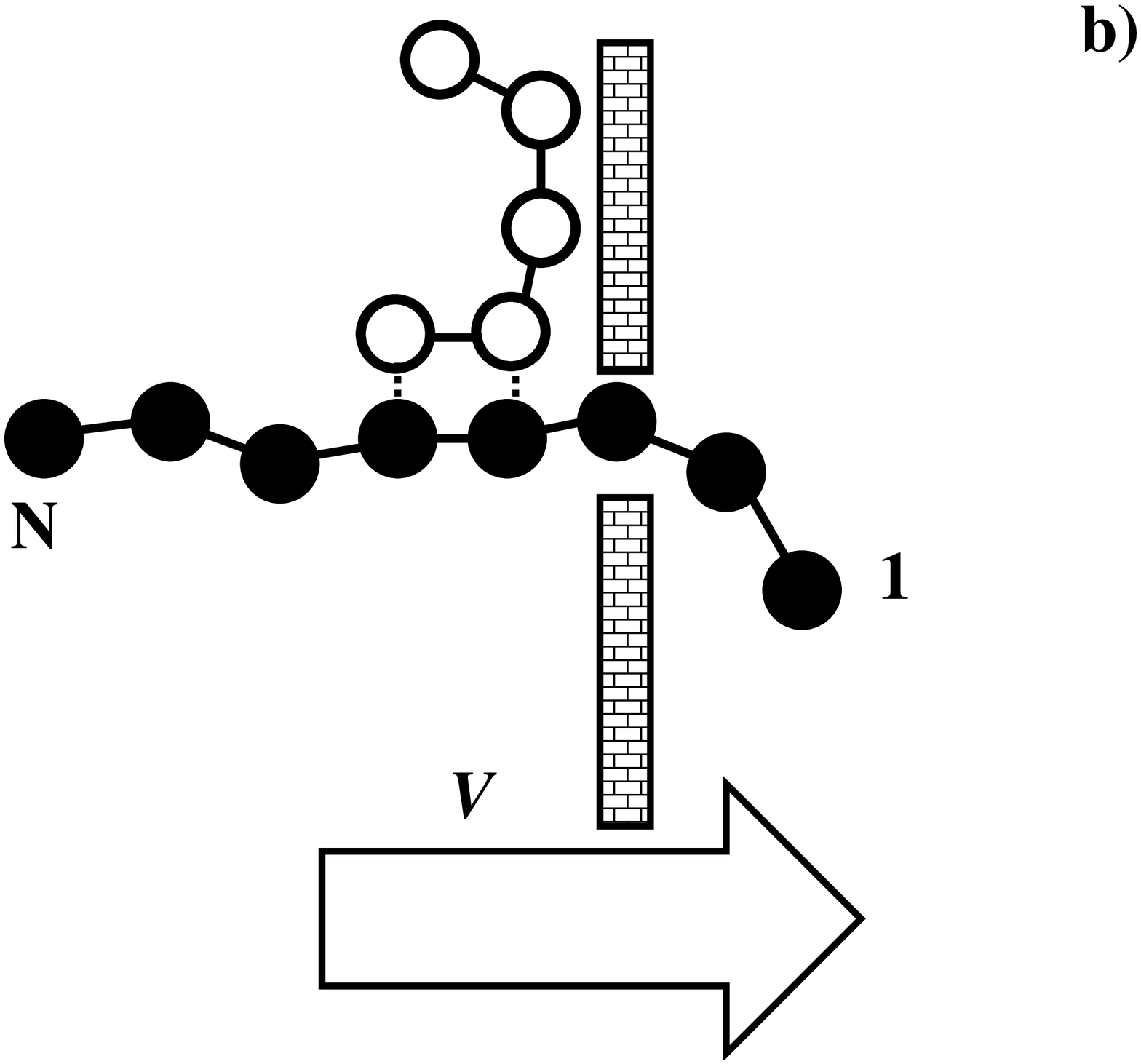}}
\end{picture}
\vskip 1in
 \begin{Large} Figure 1. Kotsev and Kolomeisky \end{Large}
\end{center}
\end{figure}

\newpage

\begin{figure}[ht]
\begin{center}
\unitlength 1in
\begin{picture}(4.0,4.0)
  \resizebox{3.375in}{3.375in}{\includegraphics{Fig2.eps}}
\end{picture}
\vskip 1in
 \begin{Large} Figure 2. Kotsev and Kolomeisky \end{Large}
\end{center}
\end{figure}

\newpage

\begin{figure}[ht]
\begin{center}
\unitlength 1in
\begin{picture}(4.0,4.0)
  \resizebox{3.375in}{3.375in}{\includegraphics{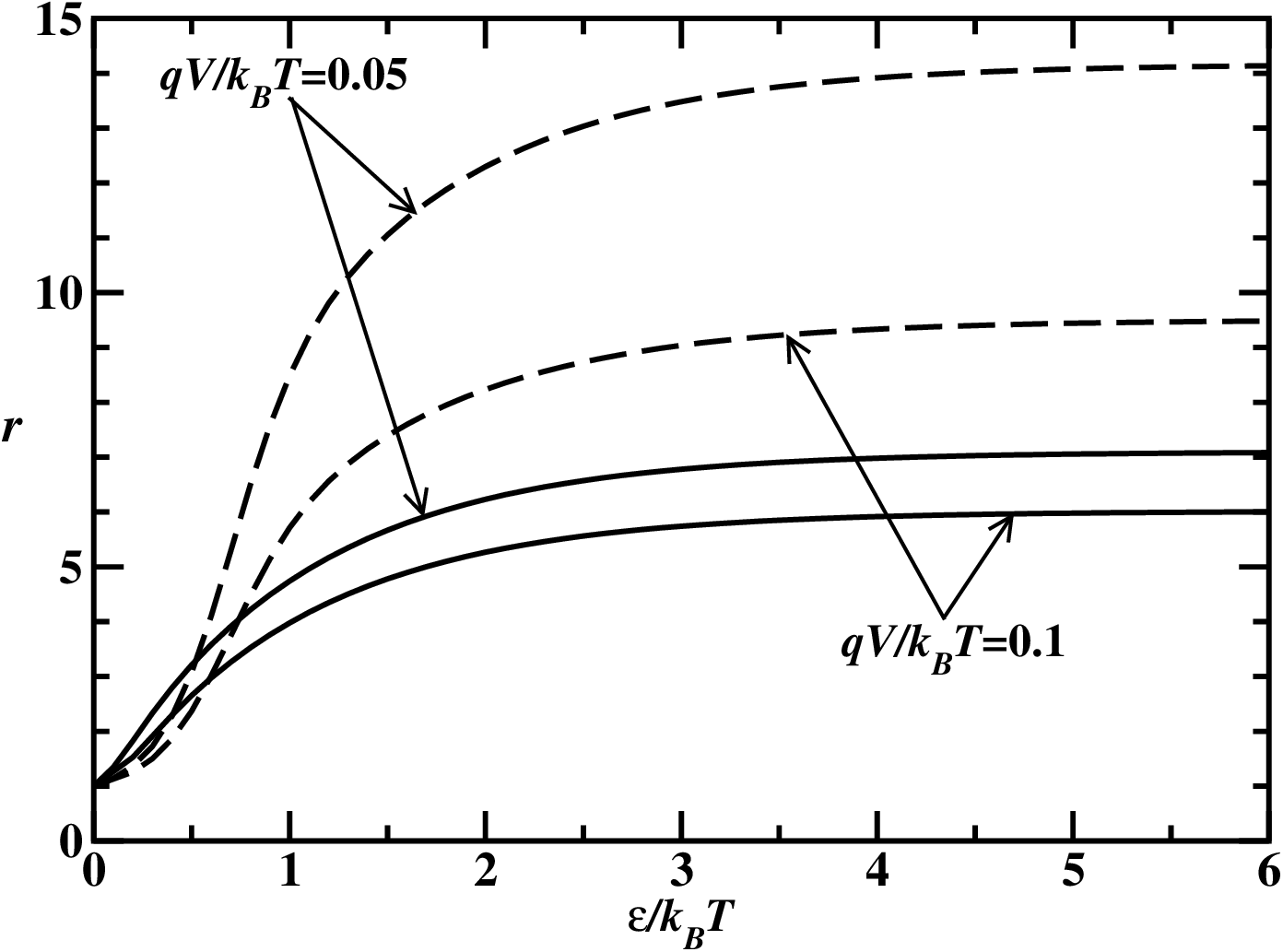}}
\end{picture}
\vskip 1in
 \begin{Large} Figure 3. Kotsev and Kolomeisky \end{Large}
\end{center}
\end{figure}

\newpage

\begin{figure}[ht]
\begin{center}
\unitlength 1in
\begin{picture}(4.0,4.0)
  \resizebox{3.375in}{3.375in}{\includegraphics{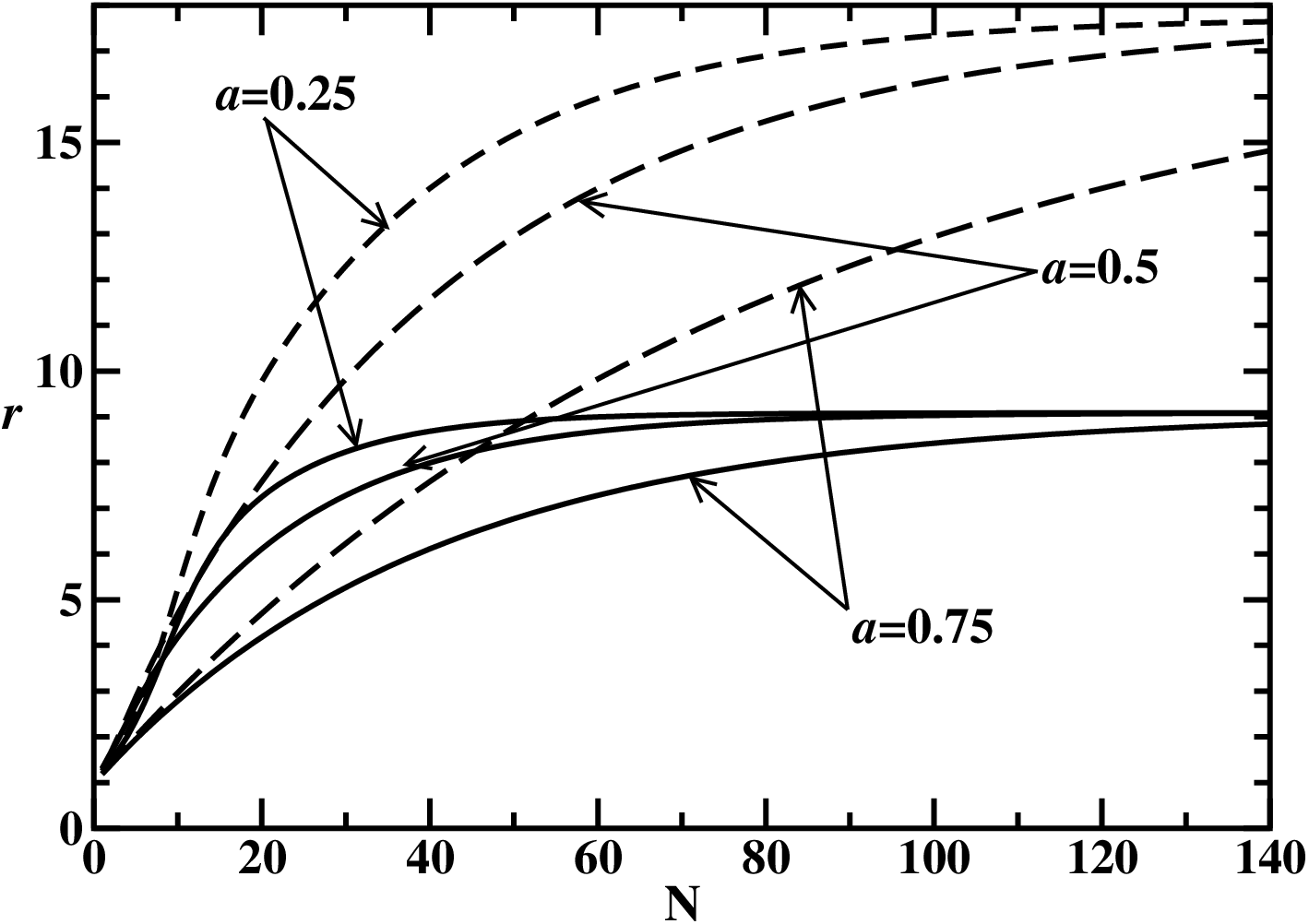}}
\end{picture}
\vskip 1in
 \begin{Large} Figure 4. Kotsev and Kolomeisky \end{Large}
\end{center}
\end{figure}

\newpage

\begin{figure}[ht]
\begin{center}
\unitlength 1in
\begin{picture}(4.0,4.0)
  \resizebox{3.375in}{3.375in}{\includegraphics{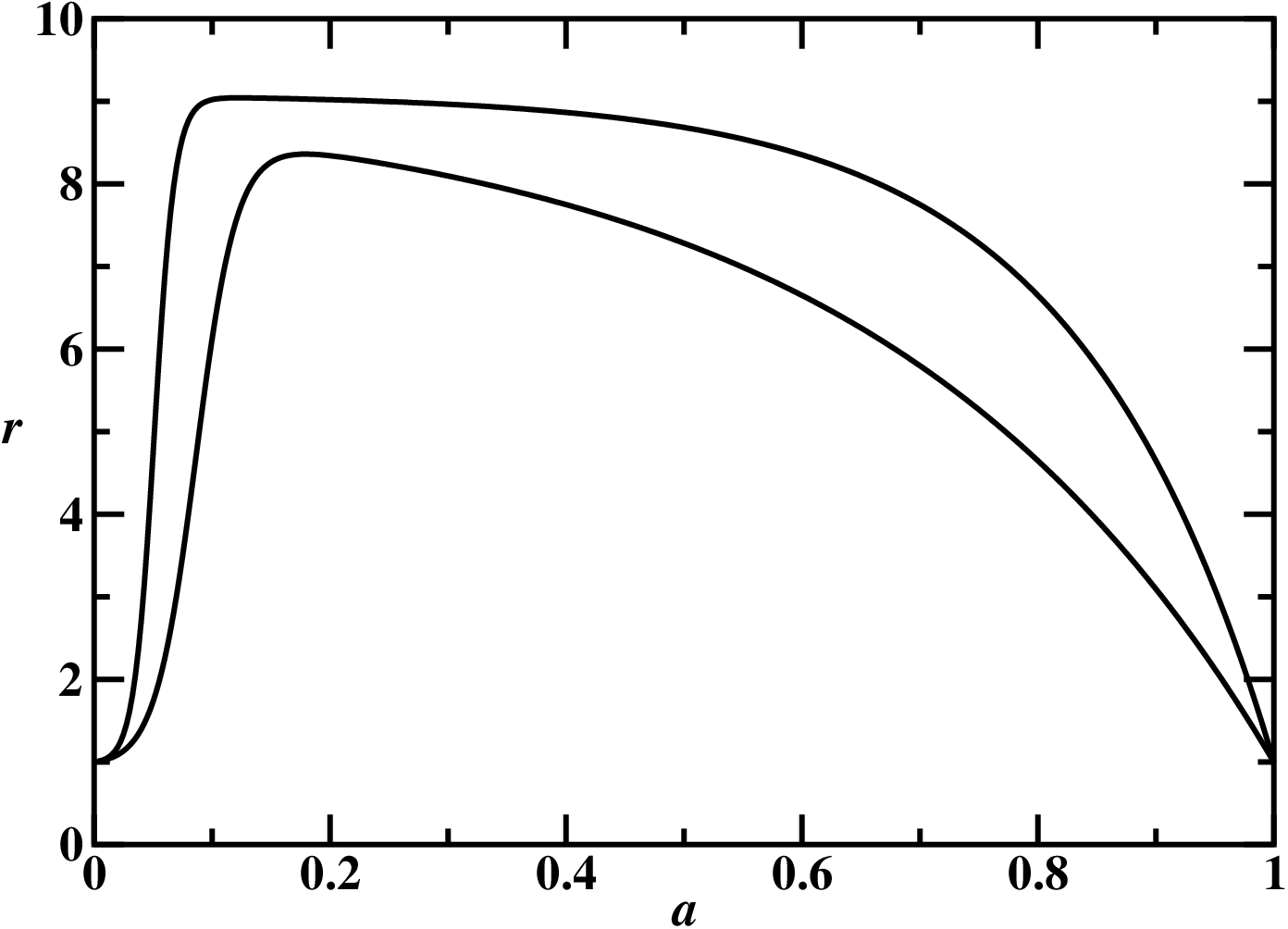}}
\end{picture}
\vskip 1in
 \begin{Large} Figure 5. Kotsev and Kolomeisky \end{Large}
\end{center}
\end{figure}

\end{document}